\documentclass[runningheads]{llncs}
\usepackage[T1]{fontenc}

\usepackage{amsmath}
\usepackage{amssymb}
\usepackage{multirow}
\usepackage{graphicx}

\begin{document}
\title{ASD-Diffusion: Anomalous
Sound Detection with Diffusion Models}
%
%
\author{Fengrun Zhang\inst{1}\orcidID{0009-0005-9488-5077} \and
Xiang Xie\thanks{Corresponding author}\inst{1}\orcidID{0000-0002-5135-0110} \and
Kai Guo\inst{1}\orcidID{0009-0003-3504-0571}}
%
\authorrunning{F. Zhang et al.}
%
\institute{Beijing Institute of Technology, Beijing 100081, China\\
\email{\{zhangfengrun, xiexiang, guokai\}@bit.edu.cn}\\
\url{https://www.bit.edu.cn} }
\maketitle              
\begin{abstract}
Unsupervised Anomalous Sound Detection~(ASD) aims to design a generalizable method that can be used to detect anomalies when only normal sounds are given. In this paper, Anomalous Sound Detection based on Diffusion Models (ASD-Diffusion) is proposed for ASD in real-world factories.
 In our pipeline, 
 the anomalies in acoustic features are reconstructed from their noisy corrupted features into their approximate normal pattern. Secondly, a post-processing anomalies filter algorithm is proposed to detect anomalies that exhibit significant deviation from the original input after reconstruction. Furthermore, denoising diffusion implicit model is introduced to accelerate the inference speed by a longer sampling interval of the denoising process. The proposed method is innovative in the application of diffusion models as a new scheme. Experimental results on the development set of DCASE 2023 challenge task 2 outperform the baseline by 7.75\%, demonstrating the effectiveness of the proposed method.

\end{abstract}
\section{Introduction}
The purpose of anomalous sound detection (ASD) in the industrial scene is to monitor the machine's condition by distinguishing between normal and anomalous machine-generated sounds. Detection and classification of acoustic scenes and events (DCASE) challenge and workshop is committed to advancing the field of sound event detection. Since 2020, ASD has been adopted as a new task and held every year until now in DCASE challenge~\cite{Koizumi2020}.
Previous methods achieve better performance by using test data in development set to tune hyper-parameters of the model~\cite{DCASE2023baseline2023}.
However, in some practical conditions, due to the diversity of operational conditions and atypical anomalies, it is challenging to collect anomalous sounds with comprehensive pattern coverage and collect anomalous data for tuning. 

Considering all the above factors, a main goal of DCASE 2023 task 2 is to detect anomalous sounds when only normal sounds are given without tunable hyper-parameters of the trained model for each machine type, which is called first-shot ASD.

A series of methods, which can be generally divided into self-supervised and unsupervised methods, have been proposed to tackle these issues. 
Self-supervised ASD introduces classification as an auxiliary task to calculate anomalous degree in accordance with classification confidence. 
However, since classification-based self-supervised approaches extremely rely on additional labels (i.e. machine ID or attribute) from metadata \cite{almudevar23_interspeech,9746207,10096054}, effectiveness may degrade when auxiliary labels are limited or domain shifts occur~\cite{DCASE2023baseline2023}.

Unsupervised ASD approaches minimize the negative log-likelihood or reconstruction error as the optimization objective and learn the distribution only from the acoustic features of normal sounds. Anomalies of audio are detected by the inner likelihood of the learned distribution or the reconstruction error of generated samples. A flurry of generative models have been previously explored in ASD, such as variational autoencoder (VAE)~\cite{Daniluk2020}, generative adversarial network (GAN)~\cite{10096813}, and normalizing flows (NF)~\cite{9414662}. 
In recent studies, denoising diffusion probabilistic model (DDPM)~\cite{ho2020denoising}, as an emerging generative model, has attracted much attention from researchers in many fields. It has been proven that DDPMs are capable of generating samples from complex data distributions with broader pattern coverage than VAEs and GANs~\cite{NEURIPS2021_49ad23d1}. These properties are considered suitable for anomaly detection that lacks anomalous samples. Recent advances in computer vision also indicate that DDPM is well suited to anomaly detection tasks. Until now, DDPM has been used for anomaly detection in images. AnoDDPM~\cite{DBLP:conf/cvpr/WyattLSW22} achieves a huge improvement over GAN-based approaches in medical image anomaly detection. DiffusionAD~\cite{zhang2023diffusionad} outperforms other methods in general image anomaly detection. However, 
applying diffusion models to ASD remains challenging and has not been explored. 

Since the high-dimensional time-frequency information in audio can be intuitively represented in the acoustic features~(i.e. mel-spectrogram), employing diffusion models for anomaly detection in these acoustic features is a reasonable choice. Inspired by the works mentioned above, we propose ASD-Diffusion, a novel diffusion-based ASD approach. The main contributions of this paper can be summarized as follows:
\begin{itemize}
\item A diffusion-based approach to ASD. To the best of our knowledge, ASD-Diffusion is the first time that diffusion models have been applied to the field of ASD. 
\item A carefully designed post-processing anomalies filter~(AF) algorithm, which is well suited for anomaly detection in samples reconstructed by diffusion models. Meanwhile, it can also be used for anomaly localization.
\item For problems of long sampling timesteps existing in DDPM, we introduce denoising diffusion implicit model (DDIM)~\cite{song2020denoising} in the inference process to accelerate sampling. 
\end{itemize}

\section{Methods}
\subsection{Diffusion Models for ASD-Diffusion}
\subsubsection{DDPM}
\begin{figure}[h]
  \centering
  \includegraphics[width=0.9\linewidth]{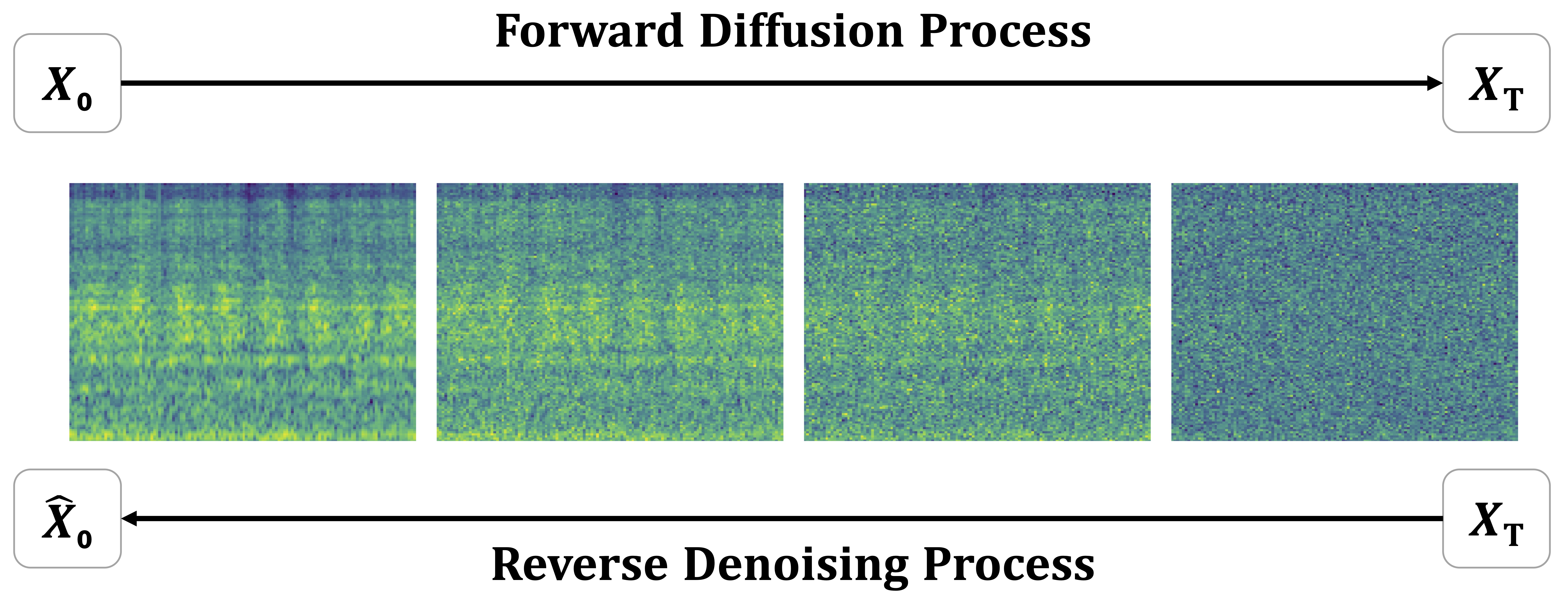}
  \caption{Forward and reverse process of DDPM.}
  \label{fig:DDPM}
\end{figure}
In general, DDPM specifies a forward diffusion process and a reverse denoising process illustrated in Fig.~\ref{fig:DDPM}. The input data are gradually disturbed by adding Gaussian noise for a few timesteps in the forward diffusion process and DDPM is guided to reconstruct target noise-free data from corrupted data in the reverse denoising process. Assume that the distribution of normal sounds is $\phi(x)$, the forward diffusion process is defined as
\begin{equation}
\label{equ:forward}
q\left(x_{t} \mid x_{0}\right)=\mathcal{N}\left(x_{t} \mid x_{0} \sqrt{\bar{\alpha}_{t}},\left(1-\bar{\alpha}_{t}\right) \mathbf{I}\right)
\end{equation}
\begin{equation}
\label{equ:xt}
x_{t}=x_{0} \sqrt{\bar{\alpha}_{t}}+\epsilon_{t} \sqrt{1-\bar{\alpha}_{t}}, \quad \epsilon_{t} \sim \mathcal{N}(0, \mathbf{I})
\end{equation}
where data $x_{0}\sim\phi(x)$ is transformed into noisy data $x_{t}$ for $t \in\{0,1, \ldots, T\}$ by adding noise for $t$ timesteps to $x_{0}$. Here, $\bar{\alpha}_t=\prod_{i=0}^t \alpha_i=\prod_{i=0}^t\left(1-\beta_i\right)$ and $\beta_i \in(0,1)$ represents the noise variance schedule.
This can be defined as a schedule from $\beta_1=10^{-4}$ to $\beta_T=10^{-2}$~\cite{ho2020denoising,DBLP:conf/icml/NicholD21,jabri2022scalable}.

 As $x_T$ is shown in Fig.~\ref{fig:DDPM}, the distribution of $x_{0} $ is gradually disrupted and approaches Gaussian noise when t increases. 
A neural network $\epsilon_\theta\left(x_t, t\right)$ is trained to predict added noise $\epsilon$ by minimizing the training objective with mean squared error~(MSE) loss:

\begin{equation}
\label{equ:loss}
\mathcal{L}=E_{t\sim[1-T],x_0\sim q(x_0),\epsilon\sim\mathcal{N}(0,\mathbf{I})}
\left(\begin{Vmatrix} \epsilon-\epsilon_\theta\left(x_t,t\right)\end{Vmatrix}^2\right)
\end{equation}
At inference phase, $x_{t-1}$ is reconstructed from previous step $x_{t}$ in reverse process with the diffusion model $\epsilon_\theta\left(x_t, t\right)$ according to:
\begin{equation}
\label{equ:reverse}
x_{t-1}=\frac{1}{\sqrt{\alpha_t}}\left(x_t-\frac{1-\alpha_t}{\sqrt{1-\bar{\alpha}_t}} \epsilon_\theta\left(x_t, t\right)\right)+\tilde{\beta}_t z
\end{equation}
in which, $z \sim \mathcal{N}(0, \mathbf{I})$ and $\tilde{\beta}_{t}=\frac{1-\bar{\alpha}_{t-1}}{1-\bar{\alpha}_{t}}\beta_{t}$. $x_{0}$ is reconstructed to fit $\phi(x)$ from $x_{t}$ in a way of Markovian chain. 

\subsubsection{DDIM}
DDIM is generalized from DDPM via a class of non-Markovian diffusion processes. In DDIM, sample $x_{t-1}$ can be generated from sample $x_{t}$ via:
\begin{align}
    \begin{aligned}
\label{equ:ddim}
x_{t-1} & =\underbrace{\sqrt{\bar{\alpha}_{t-1}}\left(\frac{x_{t}-\sqrt{1-\bar{\alpha}_{t}}\epsilon_{\theta}^{(t)}(x_{t})} 
{\sqrt{\bar{\alpha}_{t}}}\right)}_{\text{“predicted }\boldsymbol{x}_{0}\text{”}} + \\
&\underbrace { \sqrt { 1 - \bar{\alpha} _ { t - 1 }-\sigma_{t}^{2}}\cdot\epsilon_{\theta}^{(t)}(x_{t})}_{\text{“direction pointing to }\boldsymbol{x}_{t}"}+\underbrace{\sigma_{t}\epsilon_{t}}_{\text{random noise}}
\end{aligned}
\end{align}
in which, $\epsilon_{t}\sim\mathcal{N}(0,I)$ is standard Gaussian noise independent of $x_{t}$. 

Different values of $\sigma_{t}$ will lead to different generative processes. When \begin{footnotesize}$\sigma_{t}=\sqrt{(1-\bar{\alpha}_{t-1})/(1-\bar{\alpha}_{t})}\sqrt{1-\bar{\alpha}_{t}/\bar{\alpha}_{t-1}}$\end{footnotesize} for all $t$, the forward diffusion process becomes Markovian, and the generative process becomes a DDPM. When $\sigma_{t} = 0 $ for all $t$, samples are generated from latent variables with a fixed procedure (from $x_{T}$ to $x_{0}$), the process becomes DDIM. This fixed denoising procedure results in a more stable reconstruction to approach $\phi(x)$.
The specific derivation process of Eq.\ref{equ:ddim} can be seen in~\cite{song2020denoising}.

In DDIM, since the forward and reverse processes are non-Markov, samples can be reconstructed with a larger sampling interval in the reverse process, saving a lot of computing resources. Meantime, the training objective is also MSE loss shown in Eq.\ref{equ:loss}, which means that there is no difference in the training process with DDPM.

\subsection{Anomaly Detection with Diffusion Models}
\begin{figure}
  \centering
  \includegraphics[width=0.8\textwidth]{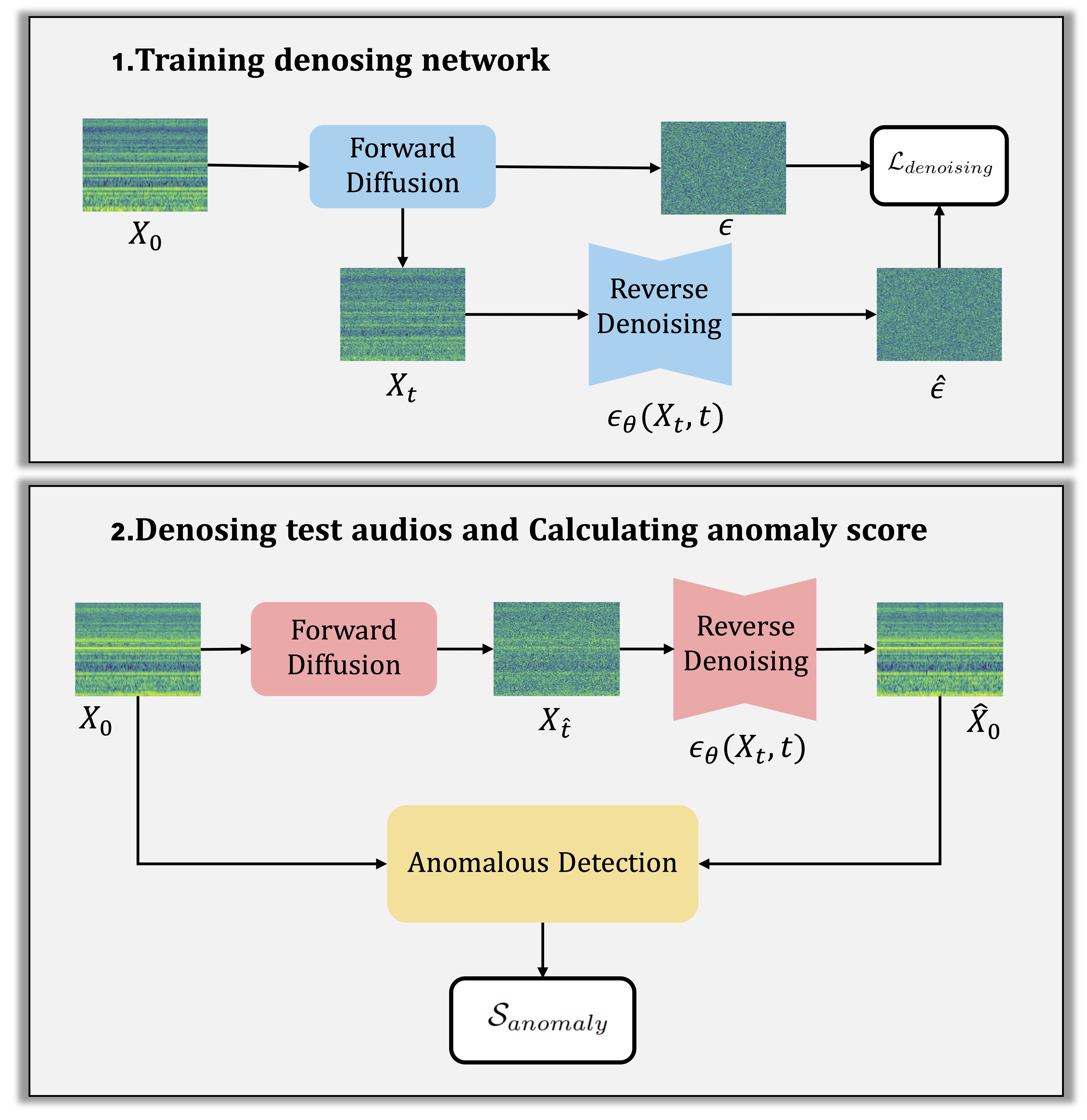}
  \caption{The overview of ASD-Diffusion. In stage 1, $x_{t}$ is obtained by adding noise $\epsilon$ to $x_{0}$ through forward diffusion. A neural network $\epsilon_\theta\left(x_t, t\right)$ is trained to estimate the noise $\hat{\epsilon}$ from $x_{t}$. In stage 2, $\epsilon_\theta\left(x_t, t\right)$ reconstructs $\hat{x}_{0}$ on $x_{\hat{t}}$ after forward diffusion, $\mathcal{S}_{anomaly}$ is then calculated by an anomaly detection function.}
  \label{fig:overall}
\end{figure}

 The overall architecture of ASD-Diffusion is illustrated in Fig.~\ref{fig:overall}.
In our work, filterbank~(FBank) features extracted from waveform is chosen for anomaly detection. Since the difference between anomaly and normality can be roughly divided into frequency domain and time domain, FBank is considered suitable for anomaly detection that contains abundant time-frequency information.

During the training stage, ASD-Diffusion corrupts the FBank of normal samples $x_{0}$ to $x_{t}$ by adding Gaussian noise with a random parameter $t \in\{0,1, \ldots, T\}$, and noise scale is controlled by $\alpha_{t}$ in Eq.\ref{equ:forward}. Then the denoising network $\epsilon_\theta\left(x_t, t\right)$ predicts the added noise of $x_{t}$. The denoising loss in Eq.\ref{equ:loss} can be simplified as:
\begin{equation}
\label{equ: simple loss}
\mathcal{L}_{denoising}=\begin{Vmatrix}\epsilon_{t}-\epsilon_{\theta}\left(x_{t},t\right)\end{Vmatrix}^{2}
\end{equation}
where $\epsilon_\theta\left(x_t, t\right)$ learns the distribution of normal samples through minimizing $\mathcal{L}_{denoising}$. 

During inference, since anomalous samples share distributions different from $\phi(x)$, an effective method is to corrupt the anomalous samples by forward diffusion and reconstruct them into their approximate normal samples in $\phi(x)$. Then the anomalies are detected by comparison between original and reconstructed samples.
In our method, a strategy of partial diffusion is adopted. The query samples are firstly corrupted with Gaussian noise with a fixed parameter $\hat{t}$. The hyper-parameter $\hat{t}$ is set to cause the anomalous regions indistinguishable from normal and retain some characteristics of the energy distribution instead of destroying the FBank into Gaussian noise totally~\cite{li2023your}. Finally, the corrupted samples are reconstructed within $\phi(x)$. 

Anomaly detection is achieved by comparing the difference between the reconstructed and query samples. The widely used mean absolute error~(MAE) calculates anomaly scores from the whole FBank, where the environmental noise is also used for calculation. Since both original and reconstructed samples contain pixel-level noise interference, which is considered redundant for anomaly detection, an AF algorithm is proposed to achieve better anomaly detection. In the experiment, we find that compared with normal sounds, most anomalous patterns appear at inappropriate frequencies, which are reflected in FBank as unreasonable energy areas with some locations. The anomaly regions are often a small part of the whole features. So we introduce AF for post-processing.
The AF function proposed can filter out regions with little differece between the reconstructed samples and the original samples. 
 The anomaly score $\mathcal{S}_{anomaly}$ is calculated via:
\begin{equation} 
\label{equ: simple loss}
S_{anomaly} = \frac{1}{FT} \sum_{\text{Topk}} \text{Relu}(x_{ij} - \hat{x}_{ij})
\end{equation}
where $x$ represents the original sample and $\hat{x}$ represents the reconstructed sample. $T$ is the number of frames in the FBank and $F$ is the number of the mel filters. The AF functions are a simple ReLU with TopK function or just a TopK function.
In real practice, due to the diversity of anomalies from different machine types, we adopt multiple AF functions for different machines to get the best performance and verify the upper limit of the algorithm. Since audio commonly exhibits continuity in both spectral and temporal domains,  the AF function can well filter out the anomalous regions in multiple domains by setting the appropriate K, which represents the top k largest data values.  

\section{Experiments}
\label{sec:experiment}
\subsection{Dataset}
The experiments are carried out on the DCASE 2023 task 2 development dataset (conducted on ToyADMOS2~\cite{Harada2021} and MIMII DG~\cite{Dohi2022}) including seven machine types. Each machine type in the dataset has one section that contains data for training and testing.  Each audio recording is single-channel with a duration of 6 to 18 sec and a sampling rate of 16 kHz.

Domain shift is introduced to reflect changes in the working conditions of machines. Most of the training data comes from the source domain. Each section of a machine type contains: (a) 990 clips of normal sounds in the source domain for training. (b) 10 clips of normal sounds in the target domain for training. (c) 100 clips each of normal and anomalous sounds including data from both domains for testing.
\vspace{-0.4cm}

\subsection{Experimental Settings}
\begin{table}
\vspace{-0.4cm}
  \begin{center}
    \caption{Hyper-parameters of diffusion, denoising U-net and training process.}
    \label{hyp}
    \resizebox{50mm}{!}{
    \begin{tabular}{c|c} 
      \hline
      Forward timestep T  & 1000  \\
      Reverse timestep $\hat{t}$ &  280  \\
      Noise schedule  & sigmoid~\cite{jabri2022scalable}  \\
      DDIM sampling interval &  4  \\
      \hline
      Channels &  64  \\
      Channels multiple &  (1, 2, 4, 8) \\
      Head &  4  \\
      Attention resolutions &  32  \\
      \hline
      Optimizer & Adam~\cite{kingma2014adam} \\
      Learning rate & 1e-4 \\
      EMA rate & 0.995 \\
      Training steps & 64000 \\
      Batch size & 24 \\
      \hline 
    \end{tabular}}
  \end{center}
\end{table}
The 128-dimensional FBank is extracted on 25ms hann windows with 10ms shifts after 1024-point Fast Fourier Transformation (FFT), then the magnitude is normalized to $ [0,1]$. The FBank of each audio is divided into multiple segments of 128 × 128 by applying a sliding window. Since diffusion models do not require a large amount of data, the hop size of the sliding window is 128 for training and 5 for testing. 
The model is trained on a single NVIDIA 3090 GPU and implemented with PyTorch.
As for training parameters, the U-net architecture is adopted as the denoising network. The hyper-parameters are listed in Table.~\ref{hyp}. Reverse timestep $\hat{t}$ is chosen by experience, since the data is fully corrupted with larger  $\hat{t}$, the reconstruction error may not be an effective detector. Similarly, if the corruption is minimal with smaller  $\hat{t}$, it will also be useless.
\subsection{Evaluation Metrics}
Area under receiver operator characteristic curve~(AUC) is the most widely used metric in ASD, since anomaly detection is essentially a binary classification task. Same to the DCASE 2023 challenge, we adopt source-AUC~(sAUC), target-AUC~(tAUC) and partial-AUC~(pAUC) as the evaluation metrics. Then the final system score is obtained by calculating the harmonic mean~(hmean) for all machine domains and types. Compared with arithmetic mean, hmean is more susceptible to the influence of low values, which is adopted to evaluate the overall performance of ASD systems~\cite{Harada_arXiv2023_01}.

\section{Results}
\label{sec:result}
\subsection{Main Results}

To demonstrate the effectiveness of ASD-Diffusion, other unsupervised methods are chosen for comparison. For fairness, we first compared all unsupervised methods from the top five teams in DCASE 2023, specifically those not employing machine ID or machine attribute. AE (MAHALA)~\cite{Harada_arXiv2023_01} is the baseline provided by the challenge organizers. GAN-VAE~\cite{BaiJLESS2023} is the fourth team in the challenge. In our method, the ReLU function is used only for bearing, fan, slider, and ToyTrain, while the parameter K in the TopK function is adjusted for each machine type  to achieve the best performance.
\begin{table}[h]
\small
\centering
\caption{Performance~(\%) comparison with unsupervised methods. Best in bold.}
\makebox[\textwidth][c]{
\resizebox{1.2\linewidth}{!}{
\begin{tabular}{l|cccccc|cccccc}
\hline
\label{results}
 \multirow{2}{*}{Machine} & \multicolumn{3}{c}{AE~(MAHALA)~\cite{Harada_arXiv2023_01}} & \multicolumn{3}{c|}{GAN-VAE~\cite{BaiJLESS2023}} & \multicolumn{3}{c}{\textbf{Ours w/o AF}} &\multicolumn{3}{c}{\textbf{Ours}}\\ 
& sAUC$\uparrow$ & tAUC$\uparrow$ & pAUC$\uparrow$ & sAUC$\uparrow$ & tAUC$\uparrow$ & pAUC$\uparrow$ & sAUC$\uparrow$ & tAUC$\uparrow$ & pAUC$\uparrow$ & sAUC$\uparrow$ & tAUC$\uparrow$ & pAUC$\uparrow$\\ \hline
bearing &  65.16 & 55.28 & 51.37 & \textbf{92.80} & \textbf{74.30} & \textbf{66.20} & 79.84 & 66.06 & 54.05 & 83.68 & 70.40 & 54.58 \\ 
fan & \textbf{87.10} & 45.98 &  59.33 & 77.00 & \textbf{73.60} & 56.20 & 77.84 & 47.40 & 60.58 & 84.10 & 59.38 & \textbf{69.05}\\ 
gearbox & \textbf{71.88} & \textbf{70.78} & 54.34 & 64.40 & 61.50 & 51.80 & 59.78 & 65.14 & 55.79 & 61.38 & 64.98 & \textbf{57.16} \\ 
slider & 84.02 & 73.29 & 54.72 & 87.80 & \textbf{78.80} & 55.10 & 90.34 & 58.46 & \textbf{61.74} & \textbf{91.98} & 61.01 & 61.68 \\ 
ToyCar  & \textbf{74.53} & 43.42 & 49.18 & 72.20 & 52.70 & 50.0 & 67.92 & \textbf{56.68} & \textbf{53.26} & 67.78 & 56.74 & 53.21 \\ 
ToyTrain & 55.98 & 42.45 & 48.13 & 61.90 & 45.80 & 48.20 & 60.32 & 54.04 & 50.47 & \textbf{63.74} & \textbf{56.30} & \textbf{52.47} \\ 
valve & 56.31 & \textbf{51.40} & \textbf{51.08} & 55.90 & 50.40 & 50.80 & \textbf{56.74} & 50.96 & 49.26 & 55.78 & 49.54 & 49.37 \\ \hline
hmean & \multicolumn{3}{c}{56.91} & \multicolumn{3}{c|}{60.69} & \multicolumn{3}{c}{59.22} & \multicolumn{3}{c}{\textbf{61.32}} \\ \hline
\end{tabular}
}}
\end{table}

As illustrated in Table.~\ref{results}, ASD-Diffusion outperforms other approaches with an improvement of 7.75\% and 1.04\%, respectively, which means that our method ranks fourth on this dataset.  The overall hmean is higher than other methods, which demonstrates that ASD-Diffusion can be better generalized to more machine types. Note that the overall tAUC is substantially superior to other methods on most machine types without any domain adaptation method, even though only 10 normal audios from the target domain are provided for training. We argue that this is due to the powerful pattern coverage ability of diffusion, that is, the distribution of the target domain can be well learned without extra domain enhancements or adaptations.

In addition, we remove the proposed post-processing algorithm~(`w/o AF') for the ablation study. As can be seen in Table.~\ref{results}, the performance of ASD-Diffusion declines by 3.42\%, respectively, which demonstrates the effectiveness of AF in ASD-Diffusion.
\subsection{ Comparison with Self-supervised Methods}
In Table.~\ref{compare with self}, our method is compared with self-supervised methods among the top three teams~\cite{JiangTHUEE2023,LvHUAKONG2023,JieIESEFPT2023}. As mentioned in the introduction, self-supervised methods achieve better results due to the use of auxiliary labels for classification. In comparison, unsupervised methods are more broadly applicable and can be used even in the absence of reliable auxiliary labels. As shown in Table.~\ref{compare with self}, our method is closer to the third self-supervised method~\cite{JiangTHUEE2023}. Furthermore, we found that the main performance difference between self-supervised and unsupervised methods is the valve machines, indicating that the reconstructed-based method may not well reflect the anomalous characteristics of the valve. We consider this to be the non-stationary characters of the valve sounds~\cite{guan2023transformer} that the reconstructed acoustic features have a large deviation from the original inputs. Therefore, even the normal sounds are reconstructed poorly, showing challenges in detecting anomalies.

\begin{table}
\small
\begin{center}
\caption{Performance~(\%) comparison with self-supervised methods of the top three teams.}
\label{compare with self}  
\begin{tabular}{ l l l}
\hline
Route & Method & hmean$\uparrow$\\ \hline  
\multirow{2}{*}{Unsupervised} & GAN-VAE~\cite{BaiJLESS2023} & 60.69  \\
                              & Ours & 61.32  \\ \hline
\multirow{3}{*}{Self-supervised} & WSP-NFCDEE~\cite{JiangTHUEE2023} & 63.26 \\
                                 & Wav2Vec (ck2)~\cite{LvHUAKONG2023} & 66.56 \\
                                 & MDAM + knn~\cite{JieIESEFPT2023} & 69.25 \\
\hline
\end{tabular}
\end{center}
\vspace{-0.4cm}
\end{table}

\subsection{Visualization of Anomaly Detection}
\label{visualization}
\begin{figure}[h]
  \centering
  \includegraphics[width=\linewidth]{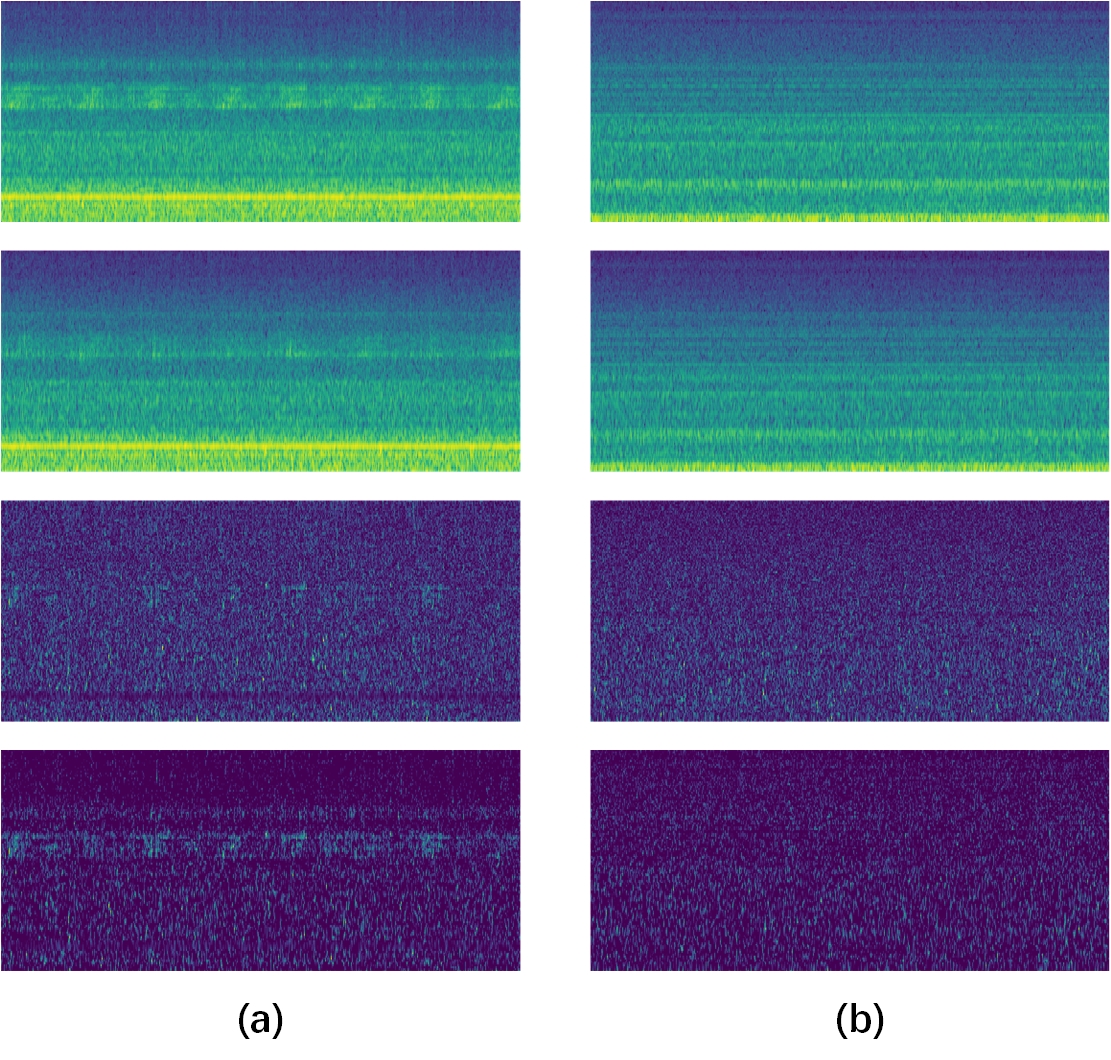}
  \caption{Visualization of an anomalous audio~(a) and a normal audio~(b). First row: original FBank. Second row: reconstructed FBank. Third row: detection result of MAE. Last row: detection result with AF.}
  \label{fig:visual}
  
\end{figure}

The results of anomalous detection can be visualized in Fig.~\ref{fig:visual}. We chose one normal and one anomalous audio from the test set of fan for comparison. The first and second rows are the original and reconstructed FBank respectively. The third and last rows are the visual detection results of MAE and the AF respectively.
In the detection results, the brighter the region, the more likely it is to contain anomalies. 
\vspace{-0.6cm}
While preserving the overall energy distribution of acoustic features, details are reconstructed at a fine-grained level. Therefore, subtle anomalies in both the time domain and the frequency domain become apparent.
From the comparison of the first and second rows in Fig.~\ref{fig:visual} (a), there is a clear difference in the middle channels of FBank, which may mean the existence of anomalies. However, subtracting and taking the absolute value causes possible anomalies to be slightly masked by background noise. The introduction of the AF function allows possible anomalies in the detection results to be retained and part of the noise to be removed.
 Whereas in Fig.~\ref{fig:visual} (b), there are virtually no conspicuous anomalous regions, which also indicates that our method can be effectively utilized for the analysis and localization of anomalies.

\subsection{Influence of AF Parameter}
\begin{figure}[h]
  \centering
  \includegraphics[width=0.9\linewidth]{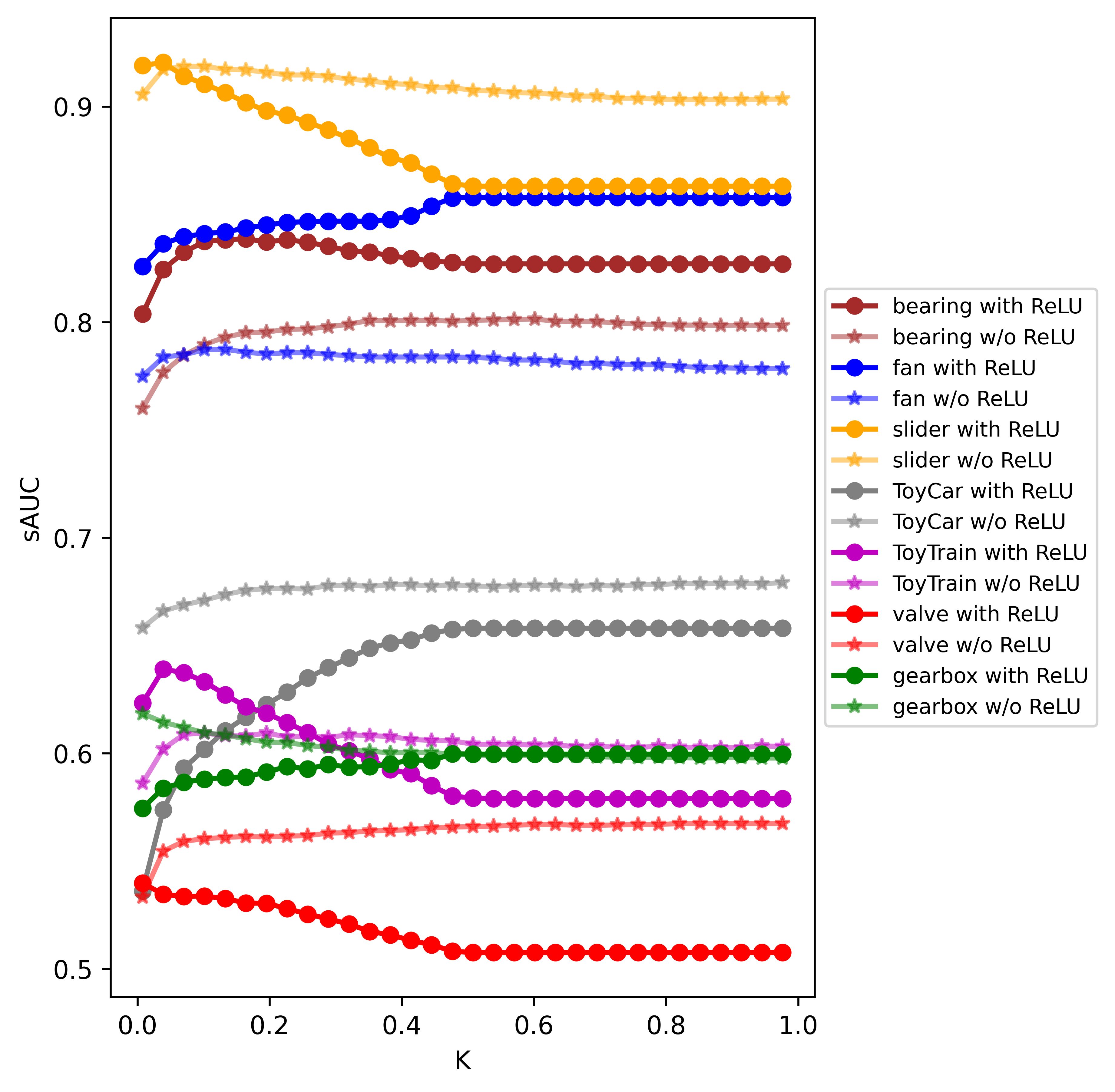}

  \caption{Performance of the proposed method under different K or ReLU functions.}
  \label{fig:topkrelu}
\end{figure}
We further explore the influence of hyperparameters in AF on the performance. We choose sAUC as the evaluation metric because it is relatively better and will not be affected by domain adaptation. In other words, different AF parameters will have a more significant impact on it. We conducted experiments from two aspects: whether to use ReLU and the percentage K of the selected TopK pixels to all pixels. In our experiments, different values of K from 0 to 1 with an interval of 0.03 are tested., while the dark- or light-colored lines represent performance with or without ReLU

As can be seen in Fig.~\ref{fig:topkrelu}, we can see that the peaks of some curves appear when K is small, which means that the abnormal locations only occupy a small part of FBank~(i.e. ToyTrain with ReLU and slider with Relu). This shows that the anomaly only occurs in a shorter time and a smaller frequency range. Besides, the ReLU function in AF has a more obvious effect on some machines~(i.e. fan, bearing and ToyTrain). This means that the anomalies of these machines are more likely to be manifested as missing high-energy regions on the FBank, as can be seen in Fig.~\ref{fig:visual} (a).

The result verifies the effectiveness of AF when facing various types of anomalies from different machines. However, tunable parameters for each machine are not possible in some scenes, such as first-shot ASD. We believe that AF is more useful for the analysis of anomalous characteristics. For example, when the ReLU function is more effective, the anomaly is more likely to be the lack of certain frequencies. To some extent, the size of K in TopK reveals the range of the anomaly.

\vspace{-0.4cm}
\subsection{Accelerating Inference Speed by DDIM}
We conduct a comparative experiment between DDPM and DDIM on the bearing. It is demonstrated in Table.~\ref{compare} that compared with DDPM, DDIM greatly improves the inference speed with lower Real Time Factor~(RTF) while maintaining better performance. Compared with DDPM, the deterministic reverse process exhibits superior consistency~\cite{song2020denoising}. For the ASD task, the consistency of the reverse process is more critical than diversity. We consider this to be a difference between generative tasks and anomalous detection tasks. 

\begin{table}
\small
\begin{center}
\caption{Performance over DDPM and DDIM.}
\label{compare} 
\begin{tabular}{l|llll}
\hline
Method & sAUC~(\%)$\uparrow$ & tAUC~(\%)$\uparrow$ & pAUC~(\%)$\uparrow$ & RTF$\downarrow$ \\ \hline
DDPM & 79.22 & 67.94 & \textbf{54.74} & 1.17 \\ 
DDIM & \textbf{83.68} & \textbf{70.40} & 54.58 & \textbf{0.29} \\
\hline
\end{tabular}
\end{center}
\end{table}

\vspace{-0.6cm}
\section{Conclusions}
In this paper, we introduce diffusion models to the field of anomalous sound detection for the first time and propose a novel method named ASD-Diffusion. Our method showcases the efficacy of diffusion models for ASD. Experimental results outperform other unsupervised methods in DCASE 2023. Meanwhile, from a practical standpoint, our method achieves interpretability and localization of anomalies with the high-quality reconstruction from DDPM. In future work, we will focus on further exploring unsupervised methods and providing better anomaly localization for ASD.

%
%
%
\bibliographystyle{splncs04}
\bibliography{samplepaper}
%







\end{document}